\documentstyle{mn}

\catcode`\"=\active\let"=\"

\def\3{\ss }

\def\c12{{1\over 2}}

\def\plusplus{\raise 0.3ex\hbox{${\scriptstyle ++}$}{}}

\newcommand{\oversim}[2]{\protect{\mbox{\lower0.5ex\vbox{%
   \baselineskip=0pt\lineskip=0.2ex
   \ialign{$\mathsurround=0pt #1\hfil##\hfil$\crcr#2\crcr\sim\crcr}}}}} 
\newcommand{\simgreat}{\mbox{$\,\mathrel{\mathpalette\oversim>}\,$}} 
\newcommand{\simless} {\mbox{$\,\mathrel{\mathpalette\oversim<}\,$}} 

\title[Dynamical friction in flattened systems: A numerical test of
Binney's approach] {Dynamical friction in flattened systems: A
numerical test of Binney's approach}
\author[J. Pe\~{n}arrubia, A. Just \&  P. Kroupa]
{Jorge Pe\~{n}arrubia$^1$
, Andreas Just$^1$, Pavel Kroupa$^{2,3}$\\
$^1$Astronomisches Rechen-Institut, M\"onchhofstrasse 12-14,
   D-69120 Heidelberg, Germany \\
$^2$Institut f\"ur Theoretische Physik und Astrophysik, Universit\"at Kiel, 
D-24098 Kiel, Germany \\
$^3$Heisenberg Fellow}

\begin{document}

\maketitle

\begin{abstract}
We carry out a set of self-consistent $N$-body calculations to
investigate how important the velocity anisotropy in non-spherical
dark-matter halos is for dynamical friction. For this purpose we allow
satellite galaxies to orbit within flattened and live dark-matter
haloes (DMHs) and compare the resulting orbit evolution with a
semi-analytic code. This code solves the equation of motion of the
same satellite orbits with mass loss and assumes the same DMH, but
either employs Chandrasekhar's dynamical friction formula, which does
not incorporate the velocity anisotropy, or Binney's description of
dynamical friction in anisotropic systems.  In the numerical and the
two semi-analytic models the satellites are given different initial
orbital inclinations and orbital eccentricities, whereas the parent
galaxy is composed of a DMH with aspect ratio $q_h=0.6$.

We find that Binney's approach successfully describes the overall
satellite decay and orbital inclination decrease for the whole set of
orbits, with an averaged discrepancy of less than 4 per cent in orbital radius
during the first 3 orbits. If Chandrasekhar's expression is used
instead, the discrepancy increases to 20 per cent. Binney's treatment
therefore appears to provide a significantly improved treatment of
dynamical friction in anisotropic systems.

The velocity anisotropy of the DMH velocity distribution function
leads to a significant decrease with time of the inclination of
non-polar satellite orbits. But at the same time it reduces the
difference in decay times between polar and coplanar orbits evident in
a flattened DMH when the anisotropic DMH velocity distribution
function is not taken into account explicitly.
Our $N-$body calculations furthermore {indicate} that polar orbits
survive about 1.6 times longer than coplanar orbits and that the
orbital eccentricity $e$ remains close to its initial value if
satellites decay slowly towards the galaxy centre. However, orbits of
rapidly decaying satellites modelled with the semi-analytic code show
a strong orbital {\em circularisation} ($\dot e<0$) not present in the
N-body computations.  

\end{abstract}

\section{Introduction}\label{sec:int}
According to current ideas galaxy formation began with small-amplitude
Gaussian fluctuations at the early stages of the Universe. In
hierarchical cosmological models, these fluctuations decrease with
increasing scales, resulting first in the formation of low-mass
objects that may merge, building up ever more massive structures.  The
shape and morphology of these objects are strongly dependent on the
cosmological models, as one can conclude from N-body computations,
although none of them predict spherical structures.  The most
successful hierarchical theory is the Cold Dark Matter model (CDM). In
this framework, aspherical bound dark matter haloes (DMHs) form as a
result of gravitational clustering.  The inclusion of gas dynamics in the CDM simulations (Udry \& Martinet 1994, Dubinsky 1994) results to a Gaussian distribution of DMH {density aspect ratios},
$q_h \equiv c/a >0$, where $c$ and $a$ are the minor and major axes of
an oblate spheroid, of mean $<q_h>=1/2$ and dispersion equal to~0.15 (Dubinsky 1994).
The degree of asphericity depends on the dark matter nature. For
example, numerical computations based on the $\Lambda$CDM predict halo axis-ratios of $q_h\simeq 0.7$ (Bullock 2001),
Hot Dark Matter models lead to haloes as round as $q_h=0.8$
(Peebles 1993), whereas candidates such as cold molecular gas
(Pfenniger, Combes \& Martinet 1994) and massive decaying neutrinos
(Sciama 1990) may produce halo profiles as flattened as $q_h=0.2$.
 
Observationally, measuring galaxy axis-ratio is complicated: (i)
Stellar kinematics: Olling \& Merrifield (2000) obtain an axis-ratio
of $q_h\approx0.8$ for our Galaxy.  This method has the disadvantage
of having access to information of our Galaxy only on small 
scales. (ii) The flying gas layer method (Olling 1996, Becquaert,
Combes \& Viallefond 1997): assuming that the HI emission comes from
gas in hydrostatic equilibrium in galactic potentials axis-ratios as
low as $q_h\approx0.3$ are obtained for the galaxies NGC 891 and 4244,
(iii) Warping gas layer: Hofner \& Sparke (1994) find axis-ratios of
approximately 0.7 for NGC 2903 and of $q_h\approx0.9$ for NGC 2841,
3198, 4565 and 4013, (iv) X-ray isophotes: Boute \& Canizares (1998)
measure values of $q_h\approx0.5$ for NGC 3923, 1332 and 720, (v)
Polar ring galaxies: Arnaboldi et al. (1993) and Sackett et al. (1994)
find an axis-ratio of $q_h\approx0.3$ for NGC 4650A, 0.5 for the
galaxy A0136-0801 and 0.6 for AM2020-504, (vi) Precessing dusty discs:
Steinman-Cameron, Kormendy \& Durisen (1992) measure an axis-ratio of
0.9 for the galaxy NGC 4753.

Another method, which we focus on here, is the analysis of satellite
dynamics. There are two main different approaches to infer the halo
shape from satellites. First, one may attempt to reproduce the
observed tidal streams of Milky Way satellites as done, for instance,
by Ibata et al. (2001) and Law et al. (2003) who, from measurements of velocity, position
and structure of the Sagittarius dwarf galaxy, constrain the initial
parameter space and, subsequently, calculate in detail the satellite
mass loss. They find that the Milky Way halo potential cannot be more flattened than
$\approx 0.8$, otherwise tidal streams would be too spread out and
thick compared to the observations due to orbital precession. The
second approach is a statistical study of satellite distribution
around spiral galaxies. Holmberg (1969), Zaritsky \& Gonz\'alez
(1999), Prada et al. (2003) and Sales \& Lambas (2003) point out that satellites around disc galaxies are found more
often aligned with the poles of the host galaxy, the so-called
'Holmberg effect', whereas Quinn \& Goodmann (1986) find in their
$N$-body study that the disc alone cannot account for the original
statistical distribution of Holmberg's data. A remedy may be sought in
the form of an extended non-spherical DMH. An anisotropic velocity
(and mass) distribution will cause a satellite's orbit to align with
the axes of the velocity ellipsoid of the host galaxy (Pe\~{n}arrubia,
Kroupa \& Boily 2001, hereinafter PKB).  In both schemes, a large
number of numerical calculations is needed. In the former approach one
should integrate several initial orbital parameters to find the best
fit to the observed satellite characteristics, whereas in the latter
approach an initial satellite sample should statistically reproduce
the distributions expected from cosmological models of DMH
formation. So far this is prohibitively time-expensive by means of any
of the present N-body algorithms.

Several studies of satellite decay have shown that, in spherical
systems, Chandrasekhar's dynamical friction (Chandrasekhar 1943) is
accurate enough if the Coulomb logarithm remains as a free parameter
to fit to the N-body data (e.g. van den Bosch et al. 1999, Colpi et
al. 1999) since it also depends on the code parameters and the number
of particles employed. For instance, Prugniel \& Combes (1992) and
Whade \& Donner (1996) find that dynamical friction is artificially
increased due to numerical noise if the particle number is
small. Semi-analytic methods that include Chandrasekhar's dynamical
friction have been demonstrated to reproduce accurately the overall
evolution of satellite galaxies (e.g Vel\'azquez \& White 1999, Taylor
\& Babul 2001) and, therefore, represent a useful tool in order to
carry out extensive studies with a large parameter space.\\
It is, however, still unclear how the inhomogeneity of the system distribution affects the satellite dynamics. Whereas Del Popolo (2003), Del Popolo \& Gambera (1998) and Maoz (1993) show that dynamical friction increases the steeper the density profile is, Just \& Pe\~narrubia (2002), following the theoretical scheme proposed  by Binney (1977), hereinafter B77, find a negligible effect on the satellite orbit due to the symmetry of the inhomogeneous terms of dynamical friction.

Although Chandrasekhar's formula, which assumes an isotropic
velocity distribution, reproduces accurately dynamical friction in
spherical systems, {an analytical study of Statler (1991) shows that in the case
of St\"ackel potentials the velocity anisotropy produce 
strong effects on the satellite orbit. 
The N-body computations of PKB confirm that Chandrasekhar's formula cannot}
account for the resulting satellite decay and evolution if the halo is
flattened. The aim of this paper is to implement a semi-analytic
scheme capable of tracking the dynamical evolution of substructures
within flattened as well as spherical DMH's.  With this purpose in
mind, we implement in our code the analytic expressions of dynamical
friction in systems with anisotropic velocity dispersions suggested by
Binney (B77) and also used by Statler, 
which reproduces
Chandrasekhar's for null anisotropy in velocity space.  We as well
compare the results of using Chandrasekhar's formula in axi-symmetric
systems to determine the effects of the velocity anisotropy on
satellite decay.

Section~\ref{sec:galsatparamflat} introduces the models. In
Section~\ref{sec:numcalflat} we provide the code and galaxy
parameters. We outline the dynamical friction approaches in
Section~\ref{sec:hdf} whereas in Section~\ref{sec:fixlam} we propose a
simple technique to fit a free parameter (in our case the Coulomb
logarithm) to the N-body data. In Section~\ref{sec:velaneff} we study
how flattened DMHs affect satellite decay and calculate the degree of
accuracy of Binney's formula.  The paper concludes with
Section~\ref{sec:concluflat}.

\section{Galaxy and satellite parameters}\label{sec:galsatparamflat}

Our DMH model is that used by PKB to facilitate an inter-comparison of
the disc and bulge effects on satellite decay (Pe\~narrubia 2003).
Here we do not add the bulge and disc components in order to distill
the dependency of dynamical friction on the velocity anisotropy in a
spheroidal DMH. 

In order to minimise computational time when constructing flattened
DMHs, we apply a highly-efficient technique using multi-pole potential
expansions to tailor the local velocity ellipsoid to the required
morphology (Boily, Kroupa \& Pe\~narrubia 2001). The {\sc MaGalie}
code scales linearly with particle number and hence we can construct
flattened DMHs consisting of $\simgreat10^6$ particles or more, in a
short computational time.

Following PKB, the flattened DMH is described by a non-singular
isothermal profile which, in cylindrical coordinates {(for $m^2[u=0]$ see eq.~\ref{eqn:m})}, can be described
as
\begin{eqnarray}
       \rho_h[m^2(0)]=\frac{M_h \alpha}{2\pi^{3/2} r_{\rm cut}}\frac{{\rm
        exp}\big[-m^2(0)/r_{\rm
       cut}^2\big]}{m^2(0)+\gamma^2}, 
\label{eqn:rho_h} \\
        \mbox{\rm with} \quad m^2(0)=R^2+z^2/q_h^2 \nonumber
\end{eqnarray}
$M_h$ being the DMH mass, $r_{\rm cut}$ the cut-off radius and
$\gamma$ the core radius, and
\begin{eqnarray}
        \alpha\equiv\{1-\sqrt{\pi}\beta{\rm exp}(\beta^2)[1-{\rm erf}
                      (\beta)]\}^{-1} =  \\ \nonumber
                  1 + \sqrt{\pi}\beta + (\pi -2) \beta^2 + O(\beta^3) 
\end{eqnarray} 
where $\beta=\gamma/r_{\rm cut}\simless 1/24$  {in our
calculations}. For $\beta = 1/24$ we find 
 $\alpha \simeq 1.076\rightarrow 1$  {already and hence thereafter we 
set $\alpha = 1$ in our analysis.}

We use self-consistent King models (King 1966) to represent our dwarf
galaxies.  These models fit early-type dwarf galaxies (Binggeli et
al. 1984).  For a comparison with the work of PKB we adopt
$c=\log_{10}(r_t/r_c)=0.8$, where $r_c$ and $r_t$ are the core and
tidal radii, respectively.

To construct the models we choose the satellite mass $M_s$ and $r_t$.
The tidal radius is determined by computing the density contrast,
$\rho_s(r_t)/\overline{\rho_g}(r_a) \sim 3$, at the apo-centric
distance ($r_a=55$~kpc) at $t=0$, $\overline{\rho_g}(r)$ being the
averaged density of the galaxy (same procedure as Vel\'azquez \& White
1999). This guarantees that all satellite particles are bound at
$t=0$.

We employ the system of units of PBK, which refers to the parameters
of the Milky Way disc. Adopting $M_d=R_d=G=1$ and according to
Bahcall, Smith \& Soneira (1982), $M_d=5.6\times10^{10}M_\odot$ and
$R_d=3.5$ kpc, the time and velocity units are, respectively,
$1.3\times10^7$ yr and $262\, \rm{km s^{-1}}$.

The values of the galaxy and the satellite parameters can be found in
Table~\ref{tab:galmods}. The parent galaxy corresponds to the model G2
of PKB, with the difference that we remove the disc and bulge
components here.

\begin{table}
\begin{tabular}{||l |l |l |l ||} \hline \hline
        & Symbol & Value(ph.u) & Value (m.u) \\ \hline  
DMH  & $N_h$ & $1\,400\,000$ & \\
(H2)\\
        & $M_h$ & $7.84 \times 10^{11} \rm{M}_\odot $ & 14.00 \\
        & $q_h$ & 0.60 & 0.60 \\
        & $\gamma$  & 3.5 kpc & 1.00\\
        & $r_{\rm{cut}}$ & 84.00 kpc & 24.00 \\ \hline
  Satellite &  $N_s$ & $40\,000$ & \\
(S1)\\
     &$M_s$ & $5.60\times 10^9 \rm{M}_{\odot}$ & 0.10 \\ 
     &$\Psi(0)/\sigma_0^2$ & 5.00  & 5.00 \\
     &$r_c$ & 0.67 kpc & 0.19 \\ 
     &$r_t$ & 7.24 kpc & 2.07 \\
     &$c$ & 1.03 & 1.03 \\ 
     &$<r>$ & 1.64 kpc & 0.47 \\     
     &$\sigma_0$ & $60.30 \rm{km s^{-1}}$ & 0.23 \\ 
\hline \hline
\newline
\newline
\end{tabular}
\caption{Primary galaxy and satellite models.  The DMH has an aspect
ratio $q_h = 0.6$. For the satellite model:
$\Psi(0)=\Phi(r_t)-\Phi(0)$, $\Phi(0)$ are the central potential and
$\Phi(r_t)$ the potential at the tidal radius (following the notation
of Binney \& Tremaine 1987); $\sigma_0$ is the velocity dispersion at
the centre, and $<r>$ the average radius of the satellite. The units
are such that {Ph.u.} means 'physical units' and {m.u.} 'model
units'. $N_h$ and $N_s$ are the number of particles used to represent
the DMH and the satellite, respectively.}
\label{tab:galmods}
\end{table}

\section{Numerical calculations}\label{sec:numcalflat}
\subsection{Code parameters}

The numerical experiments were carried out by using the code {\sc
Superbox} which is a highly efficient particle mesh-algorithm based on
a leap-frog scheme (for a detailed description see Fellhauer et al. 2000). {\sc
Superbox} {has already been} implemented in {ex}tensive studies of
satellite disruption by Kroupa (1997), Klessen \& Kroupa (1998) and
PKB.
  The program calculates the accelerations using a high order NGP (`nearest grid point') force calculation scheme based on the second derivatives of the potential. A self-consistent system of several galaxies can be treated by forming sub-grids (3D 'boxes') which follow the motion of each galaxy. Each sub-grid has three levels of resolution, the two finest levels co-move with the galaxies allowing a high-resolution calculation of the forces acting on the particles, whereas the third one covers the local 'Universe'. The finest levels are centred on the density maximum of the galaxy, which is re-computed at every time-step.

The code parameters are those of PKB. In that paper a detailed
description of the system and the grid structure is presented, whereas
here we merely give a brief description of the N-body parameters.
Our integration time step is $0.39$ Myr which is about $1/40$th the
dynamical time of our satellite. We have three resolution zones, each
with $64^3$ grid-cells: (i) The inner grid covers out to 3~radial halo
scale-lengths, providing a resolution of 350~pc per grid-cell. (ii) The
middle grid covers the whole galaxy, with an extension of 24 halo
scale-lengths (84~kpc), giving a resolution of 2.8~kpc per
grid-cell. The satellite always orbits within this grid except at the
very late stages of its evolution, avoiding cross-border effects (see
Fellhauer et al. 2000) and (iii) The outermost grid extends to 348~kpc
and contains the local universe, at a resolution of 11.6 kpc. As for
the satellite grid-structure, the resolutions are 816~pc per grid-cell
for the inner grid that extends to 24.48~kpc, 1.2~kpc per grid-cell
for the middle grid which extends to 36~kpc, and 11.6~kpc per
grid-cell for the outermost grid that covers the local universe.

The selection of grid parameters ensures the conservation of energy
and angular momentum for satellites in isolation over times as long as
our calculations to better than 1\% for all the models.

\subsection{Orbital parameters}

The parameter space of satellite galaxies is extremely large. A
complete survey should account for different satellite masses,
apo-galacticon distances, orbital eccentricities and
inclinations...etc. In this paper, we carry out a set of calculations
selecting those parameters that best reflect the effects of the
velocity dispersion anisotropy on satellite decay. These parameters
are: (i) the initial orbital inclination ($i$), defined as the angle
between the initial angular momentum vector of a satellite and the
axis perpendicular to the axi-symmetry plane (selected as the
$z$-axis). We expect orbital inclination to decrease in time as
predicted by Binney. We note that all the calculations proceed with
the same orbital sense, but this is not important since the halo is
non-rotating. (ii) The satellite's initial orbital eccentricity,
defined as $e=(r_a-r_p)/(r_a+r_p)$, where $r_a, r_p$ are the apo- and
peri-galacticon, respectively.

The  {parameters of the} numerical experiments are listed in
Table~\ref{tab:numexp}.

\begin{table}
\begin{tabular}{||l |l |l |l |l |l |r ||} \hline \hline
Name & Gal.       & Sat.       &$i$ &$e$ &$r_p$ &$r_a$ \\ 
     & model      & model      &           &    &[kpc] &[kpc] \\  \hline
H2S100 & H2 & S1 & $0^{\circ}$ & 0.5        &18  &55 \\
H2S130 & H2 & S1 & $30^{\circ}$ & 0.5     &18  &55 \\
H2S145 &  H2 & S1 & $45^{\circ}$  & 0.5     &18  &55 \\               
H2S160 & H2 & S1 & $60^{\circ}$ & 0.5     &18  &55 \\
H2S190 & H2 & S1 & $90^{\circ}$   & 0.5     &18  &55 \\ \hline
H2S100c & H2 & S1 & $0^{\circ}$  & 0.3       &30  &55 \\
H2S130c & H2 & S1 & $30^{\circ}$  & 0.3      &30  &55 \\
H2S145c & H2 & S1 & $45^{\circ}$   & 0.3     &30  &55 \\
H2S160c & H2 & S1 & $60^{\circ}$   & 0.3     &30  &55 \\
H2S190c & H2 & S1 & $90^{\circ}$   & 0.3     &30  &55 \\ \hline
H2S100e & H2 & S1 & $0^{\circ}$  & 0.7     &10  &55 \\
H2S130e & H2 & S1 & $30^{\circ}$ & 0.7     &10  &55 \\ 
H2S145e & H2 & S1 & $45^{\circ}$  & 0.7     &10  &55 \\
H2S160e & H2 & S1 & $60^{\circ}$ & 0.7     &10  &55 \\ 
H2S190e & H2 & S1 & $90^{\circ}$  & 0.7     &10  &55 \\ 
\hline \hline
\end{tabular}
\caption{The numerical experiments. The peri- and apo-galactica are    
$r_p$ and $r_a$, respectively, and $e=(r_a-r_p)/(r_a+r_p)$ is the
orbital eccentricity.} 
\label{tab:numexp}
\end{table}

\section{Halo Dynamical Friction}\label{sec:hdf}

Chandrasekhar's expression cannot explain some effects observed in
N-body calculations of satellite decay within flattened haloes
(PKB). Our aim is to check Binney's approximation (B77) for systems
with anisotropic velocity dispersion.

For simplicity, we reproduce here the analytic formul\ae~employed
throughout this study (for a detailed analysis of the friction force
see Pe\~narrubia 2003 and Just \& Pe\~narrubia 2002). \\ If the
distribution function in velocity space is axi-symmetric, the zeroth
order specific friction force is (B77)
\begin{eqnarray}\label{eqn:ffbinn}
F_{ i}=-\frac{2\sqrt{2\pi}\rho_h[m^2(0)] G^2M_s \sqrt{1-e_v^2} {\rm
ln}\Lambda}{\sigma^2_R \sigma_z}B_R v_{i}, \\ \nonumber
F_{ z}=-\frac{2\sqrt{2\pi}\rho_h[m^2(0)] G^2M_s \sqrt{1-e_v^2}{\rm
ln}\Lambda}{\sigma^2_R\sigma_z}B_z v_{z},
\end{eqnarray}
\noindent where $i=x,y$ and $(\sigma_R,\sigma_z)$ is the velocity dispersion
ellipsoid of a Schwarzschild distribution in cylindrical coordinates with constant ellipticity
$e_v^2=1-(\sigma_z/\sigma_R)^2$, ${\rm ln}\Lambda$ is the Coulomb
logarithm of the halo and
$$ B_R=\int_0^\infty dq \frac{\rm{exp}
(-\frac{v_{R}^2/2\sigma^2_R}{1+q}-\frac{v_{z}^2/2\sigma^2_{R}}{1-e_v^2+q})}
{(1+q)^2(1-e_v^2+q)^{1/2}}, $$
$$ B_z=\int_0^\infty dq \frac{\rm{exp}
(-\frac{v_{R}^2/2\sigma^2_R}{1+q}-\frac{v_{z}^2/2\sigma^2_{R}}{1-e_v^2+q})}
{(1+q)(1-e_v^2+q)^{3/2}}, $$ where $(v_R,v_z)$ are the coordinates of
the satellite velocity in this frame. 

As Binney shows, a body with mass $M_s$ will suffer a decrease of its
orbital inclination whenever $B_z>B_R$ (oblate halo). If the orbit is
either coplanar or polar, the inclination remains constant since,
respectively, the perpendicular and the planar component of ${\bf v}$
is zero. It is straight-forward to show that this expression
reproduces Chandrasekhar's for $e_v=0$,
\begin{equation}
{\bf F}_{\rm ch}=-4\pi G M_s \rho_h[m^2(0)] {\rm ln}\Lambda \big[{\rm
erf}(X)-\frac{2 X}{{\sqrt \pi}}e^{-X^2}\big]\frac{{\bf v}_s}{v_s^3},
\label{eqn:ffchan}
\end{equation}
\noindent where $X=|{\bf v}_s|/{\sqrt 2} \sigma$.

One important aspect to note is that both expressions of dynamical
friction include here an anisotropic halo density, denoted by
{$\rho_h[m^2(0)]=\rho_h[R^2+z^2/q_h^2]$} in cylindrical coordinates. 
This is the ``local
approximation'' made here but we note that the derivation of
eq.~\ref{eqn:ffchan} assumes a uniform, infinitely extended background
medium.  In practice, the local approximation implies the only
difference between both expressions to be the anisotropic terms of the
velocity distribution.

\section{The semi-analytical code}\label{sec:sac}

In order to analyse the accuracy of the analytic expressions of
dynamical friction we have constructed a semi-analytic algorithm that
solves the equation of motion of a point-mass satellite
\begin{eqnarray}\label{eqn:diffeq}
\frac{d^2{\bf x}}{d t^2}={\bf F}_g+{\bf F}_{\rm df},
\end{eqnarray}
where ${\bf F}_g$ is the specific force from the parent galaxy and
${\bf F}_{\rm df}$ that due to dynamical friction
(eqs.~\ref{eqn:ffbinn} or~\ref{eqn:ffchan}).

If the parent galaxy follows the fixed density profile of
eq.~(\ref{eqn:rho_h}), the specific force can be written in Cartesian
coordinates (Chandrasekhar 1960) as 
\begin{eqnarray}\label{eqn:f_h}
F_{g,i}=-2\pi G x_i \int^{\infty}_0
\frac{du}{(1+u)^2(1+e_h^2+u)^{1/2}}\rho_h[m^2(u)],\\ \nonumber
F_{g,z}=-2\pi G z \int^{\infty}_0
\frac{du}{(1+u)(1+e_h^2+u)^{3/2}}\rho_h[m^2(u)],
\end{eqnarray}
where $x_i=x,y$, the ellipticity of the galaxy is $e_h^2\equiv 1-q_h^2$, and
\begin{equation}
m^2(u)=\frac{R^2}{1+u}+\frac{z^2}{1-e_h^2+u}.
\label{eqn:m}
\end{equation}

The algorithm employed to solve eq.~(\ref{eqn:diffeq}) is based on the
Bulirsch-Stoer method (for a complete description see Press et
al. 1986), which provides high-accurate solutions with minimal
computational effort. This method is based on an adaptive step-size
scheme, thus, being ideal for systems with non-smooth potentials, as
may be the case for satellites on highly eccentric orbits {when disc and bulge
components are included}.

For calculating ${\bf F}_{\rm df}$ (eqs.~\ref{eqn:ffbinn} and \ref{eqn:ffchan}) the Coulomb logarithm is treated as
a free parameter to be fitted to the numerical orbits.
The
satellite mass $M_s(t)$ is obtained from the N-body data (see
Section~\ref{sec:massloss}) and is treated as an input function.
Numerical tests of orbits in a Keplerian potential show that energy and angular
momentum are conserved to $10^{-8}$ and $10^{-9}$ per orbit, respectively, after
choosing the accuracy of the Bulirsch-Stoer algorithm to be EPS$=10^{-5}$. This value
remains fixed for the calculations presented in this paper. An extensive
description of our semi-analytic code can be found in Pe\~narrubia (2003).

\section{Determining the Coulomb logarithm}\label{sec:fixlam}

The Coulomb logarithm is usually fit to numerical data with the aim of
reproducing the overall orbital evolution by means of semi-analytic
algorithms (e.g. Fellhauer et al. 2000).  This procedure may actually
be considered as a ``calibration'' of the semi-analytic code, which
must be done carefully if a detailed inter-comparison between
different schemes of dynamical friction is desired. For that reason we
present in what follows a method to describe the accuracy of the
semi-analytic scheme.

One possible way to quantify similarity of two orbits is via 
the quantity 
\begin{eqnarray}\label{eqn:Q}  
\chi^2=\frac{1}{2k}\sum^{2k}_{i=1}\big[({\bf r}_i-{\bf r}_{i,n})^2+
\sigma^2(r_0)(t_i-t_{i,n})^2\big],
\end{eqnarray}
\noindent ${\bf r}$ being the satellite position vector at the peri- and apo-galactica and $t$ the time at
which the satellite passes by these points. The subindex $n$ denotes
the numerical values and $\sigma(r_0)$ the velocity dispersion of the
DMH at the initial galacto-centre distance. The sum is over a given
number of orbits $k$. 

The definition of $\chi$ measures the divergence of the numerical
and semi-analytical satellite position. The term $\sigma \Delta t$ allows for the possibility that
the orbital periods differ during the evolution, which would lead to a secular deviation . By definition, $\chi$
is equivalent to the discrepancy between the numerical and
semi-analytical position evolution per orbit. The selection of the
maximum and minimum galacto-centre distances for comparison permits a
direct control over the orbital eccentricity evolution, although the
measure of $\chi$ can be extended to other points without loss of
generality.

The value of $k$ depends on the objectives of the study. For instance,
if the aim is to find the best calibration for a large period of time,
as it may be to reproduce the satellite decay in spiral galaxies, the $k$-value must include a number of periods large enough, so that the overall evolution of several satellite orbits can be reproduced with a single value of $\ln \Lambda$. In this paper, we limit our fit to the first satellite periods, namely, $k=3,4$, for which the differences between both approaches of dynamical friction can be clearly seen.

\begin{figure}
\vspace{8.7cm} \includegraphics{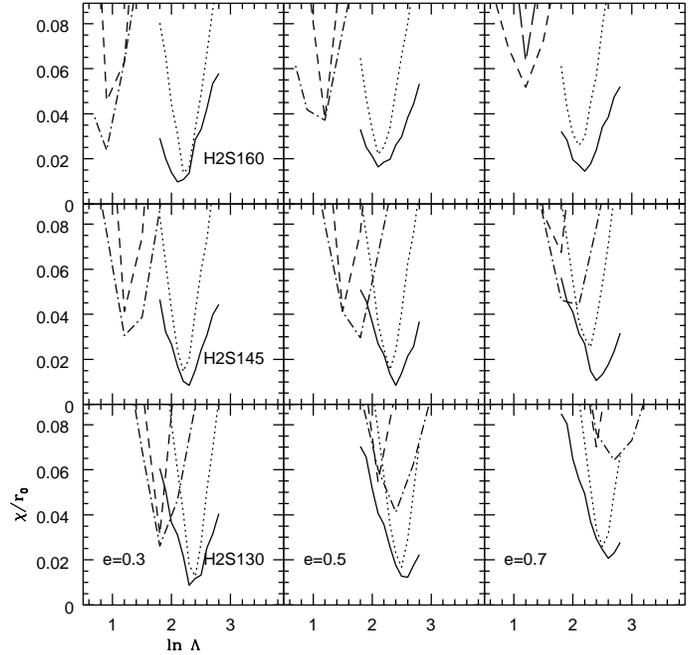}
\caption{The parameter $\chi$ for different orbital eccentricities and
inclinations. Dotted lines denote the first 4 orbits, whereas solid
lines the first 3 orbits, in both cases using Binney's expressions for
dynamical friction. Dashed and dot-dashed lines represent the results
using Chandrasekhar's expression for $k=4$ and 3 orbits,
respectively. The $\chi$ values are normalised to the initial
apo-galacticon distance $r_0=55$ kpc.}
\label{fig:box}
\end{figure}
In Fig.~\ref{fig:box} we plot $\chi$ for some of the experiments,
concretely, those with inclinations {$60^\circ$, $45^\circ$ and
$30^\circ$ (rows), with eccentricities 0.3, 0.5 and 0.7 (columns)}. For
each model, the semi-analytic code is employed to generate the
satellite orbit using Chandrasekhar's (dashed and dotted-dashed lines)
and Binney's (full and dotted lines) formula to reproduce dynamical
friction.  This figure shows that Chandrasekhar's formula poorly
describes the dependence of the satellite orbit with the initial
inclination, leading to a wider dispersion of the Coulomb logarithm
values (for this range of inclinations, between 30$^\circ$ and
60$^\circ$, ln$\Lambda\in[0.9,2.8]$). If Binney's expression is used,
the variation of $\ln\Lambda$ is highly reduced
(ln$\Lambda\in[2.3,2.5]$), which shows that this scheme provides a
much better description of the effects of anisotropic velocity
dispersion on satellite decay, independently of the orbital
inclination. The variation range of the Coulomb logarithm becomes
larger for a wider inclination spread, but $\chi$ barely presents a
dependence on the satellite eccentricity if Binney's formula is
applied. 

\begin{figure}
\vspace{9.0cm} \includegraphics{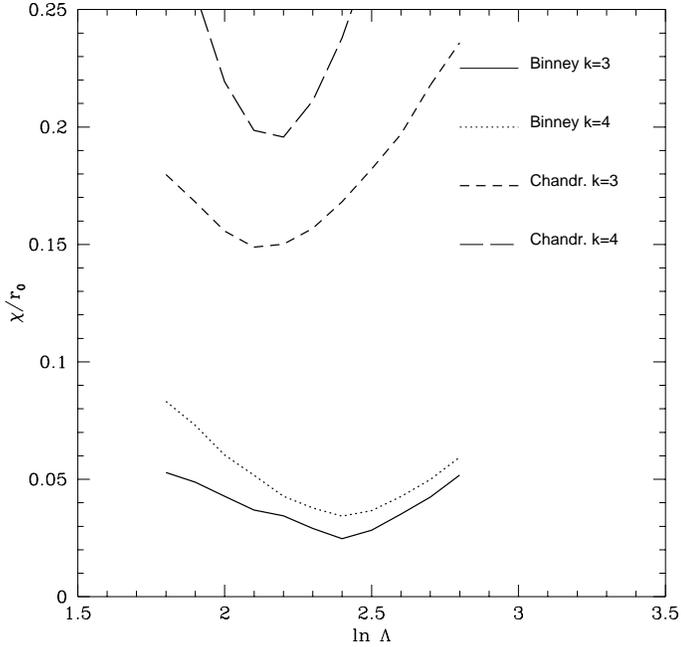}
\caption{Average of the fitting parameters over the calculations of
Table~\ref{tab:numexp}.}
\label{fig:mediaflat}
\end{figure}

If Chandrasekhar's approach is used, the Coulomb logarithm that leads
to the best fit becomes smaller as the inclination increases. Since
dynamical friction is proportional to $\ln\Lambda$, the use of the
average value implies an overestimate of the force for low
inclinations and {\em vice versa}.
The final average,
\begin{equation}
(\chi/r_0) = {1\over N_{{\rm ln}\Lambda}} \, \sum_{{\rm ln}\Lambda}
\chi_i/r_0,
\label{eq:fav}
\end{equation}
over the $N_{{\rm ln}\Lambda}$ numerical experiments of
Table~\ref{tab:numexp} is plotted in Fig.~\ref{fig:mediaflat}. This
figure shows the large discrepancies produced by Chandrasekhar's
expression if the fit is for a large range of orbital inclinations and
eccentricities, as expected. The minima of the curves determine the
values of ln$\Lambda$ that lead to the best fits, which we summarise in
Table~\ref{tab:averres}. The values of $\chi_{\rm min}$ denote the
{average error during the first $k$ orbits} associated with the fit.

For the following analysis the value of the Coulomb logarithm
implemented in our semi-analytic code will be found in
Table~\ref{tab:averres}. Looking at Fig.~\ref{fig:box}, we expect that
Chandrasekhar's friction will lead to more accurate fits for low
($i\simeq 30^\circ$) than for high inclined orbits after fixing
ln$\Lambda=2.2$.

\begin{table}
\begin{tabular}{||l  |l  |l |r |r||} \hline \hline
Friction &k& ln$\Lambda$  & $\chi_{\rm min}/r_0$ 
&$\chi_{\rm min}/(k\,r_0)$\\ \hline
Binney     & 3& 2.4 &  0.024 & 0.0080\\
                  & 4& 2.4 &  0.038 & 0.0095\\ \hline
Chandrasekhar            & 3& 2.1  & 0.147 & 0.049 \\
                  & 4& 2.2  & 0.193 & 0.048 \\
\hline \hline
\end{tabular}
\caption{Results of the fitting procedure applied to the numerical
calculations of Table~\ref{tab:numexp} for both formul\ae~of dynamical
friction. The fifth column gives relative deviation per orbit.}
\label{tab:averres}
\end{table}

\section{The velocity anisotropy effects}\label{sec:velaneff}
In this Section, we discuss in more detail various aspects of
satellite evolution.

\subsection{Satellite mass loss}\label{sec:massloss}
Tidal forces induce satellite mass loss. The satellite mass plays an
important role in determining the ultimate fate of its evolution and
survival (e.g.  Vel\'azquez \& White 1999, Klessen \& Kroupa 1998,
Johnston et al.  1999). The value of $M_s(t)$ is treated as an input
by the semi-analytical code, and is calculated using the
self-consistent {\sc Superbox} code. 

The mass remaining bound to the satellite, $M_s(t)$, is determined in
the {\sc Superbox} calculations by computing the potential energy
$\Phi_i < 0 $ of each satellite particle presumed bound to the
satellite, and its kinetic energy ($T_i$) in the satellite
frame. Following PKB, particles with 
$E_i=T_i+m_s(\Phi_i+\Phi_{{\rm ext}})>0$ 
are labelled unbound, where $m_s$
is the mass of one satellite particle (all have the same mass) and the
potentials
\begin{eqnarray}
 \Phi_i=-\sum_{i\neq j}\frac{G m_s}{\sqrt{|r_i-r_j|^2+\epsilon^2}},\\ \nonumber
\Phi_{{\rm ext}}=|\Phi_g(r_s)|,
\label{eqn:masspot}
\end{eqnarray}
the softening being $\epsilon\simeq 0.23\,$mu $=0.8\,$kpc, which is
the resolution of the inner grid focused on the satellite
centre-of-density $r_s$, and $\Phi_g$ the galaxy potential at this
point, neglecting the tidal terms. Thus, all the particles of a
satellite are assumed to feel the same external potential, which is a
useful and sufficiently accurate approximation, taking into account
that most of the bound particles are located very close to this point.

Particles with $E_i >0$ are removed and the procedure repeated until
only negative energy particles are left. As Johnston et al (1999)
show, the energy criterium permits one to distinguish those particles
that, though unbound, remain in orbits inside the tidal radius, which will escape from the satellite after some orbital periods.

The mass is calculated each $\Delta t=0.312$ Gyr, so that the
semi-analytic code interpolates the value for intermediate points at
each time-step. The error is of the order of $\Delta M(t)/\Delta t$,
going linearly with the mass loss. This means that the interpolation
might introduce not negligible differences at times where the mass
loss is significant (i.e late times of the satellite evolution).  In
this study we are not concerned in detail with the late phases of
evolution.

In the right columns of Fig.~\ref{fig:ya1}{\bf a}, {\bf b} and {\bf c}
we plot the mass evolution for different orbits. Most of the
satellites reach the inner most regions of the parent galaxy with 
a substantial fraction of their initial mass.  A comparison of these
curves with those of PKB (where a bulge and a disc component were
included) shows that the baryonic subsystems of a galaxy induce a
larger mass loss through tidal heating (e.g, Taylor \& Babul 2001,
Pe\~narrubia 2003). PKB observe in their numerical experiments that
all satellites with $M_s=0.1 M_d$ and $r_0=55$ kpc are destroyed
before the remaining bound part of the satellite reaches the central
region of the galaxy.

\subsection{Satellite decay}
One of the most important effects of dynamical friction is the
monotonic reduction of the orbital angular momentum and energy during
the satellite's evolution that leads to a progressive decrease of the
averaged galacto-centre distance. The computations carried out by PKB
show a strong dependence of the decay time on the initial inclination
that must be compared to analytic estimates.

In Fig.~\ref{fig:ya1}{\bf a}, {\bf b} and {\bf c} we plot the radius
evolution (left columns) for those models with initial $e=0.5, 0.7$
and 0.3, respectively.  From this figure, we conclude that Binney's
expression clearly produces more accurate results than Chandrasekhar's
for the whole range of orbital inclinations. This result is not
surprising due to the small dependence of the Coulomb logarithm on the
inclination and eccentricity as shown in Fig.~\ref{fig:box}.
Additionally, the value of $\ln\Lambda$ that produces the best fit for
the first few orbits also succeeds in reproducing the decay time of the
satellite.

\begin{figure}
\vspace{10.5cm}
\includegraphics{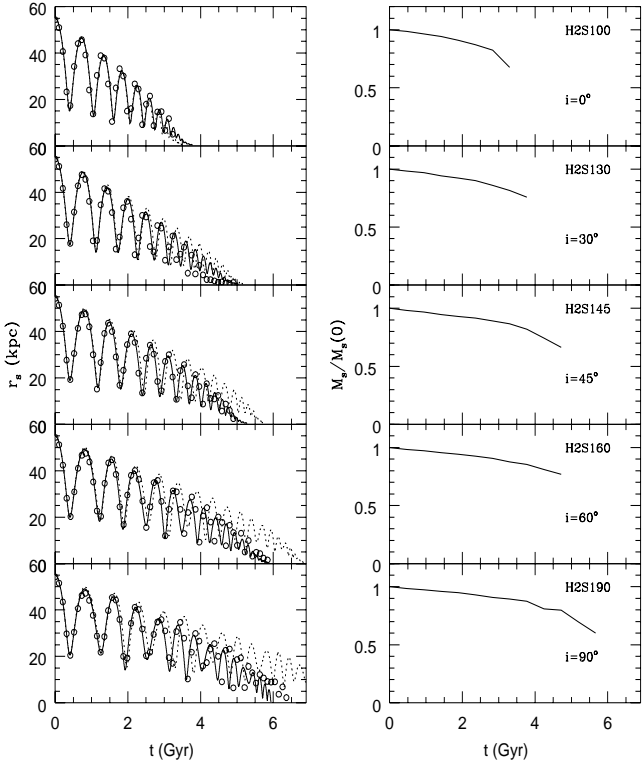}
\caption{{\bf a:} Radius and mass evolution for the models of
Table~\ref{tab:numexp} with initial $e=0.5$. Open circles denote the
numerical evolution, whereas full and dotted lines represent the data
obtained from the semi-analytic code using, respectively, Binney's
($\ln \Lambda=2.4$) and Chandrasekhar's ($\ln \Lambda=2.2$)
expressions to reproduce dynamical friction.
\label{fig:ya1}}
\end{figure}

\begin{figure}
\vspace{10.5cm} \includegraphics{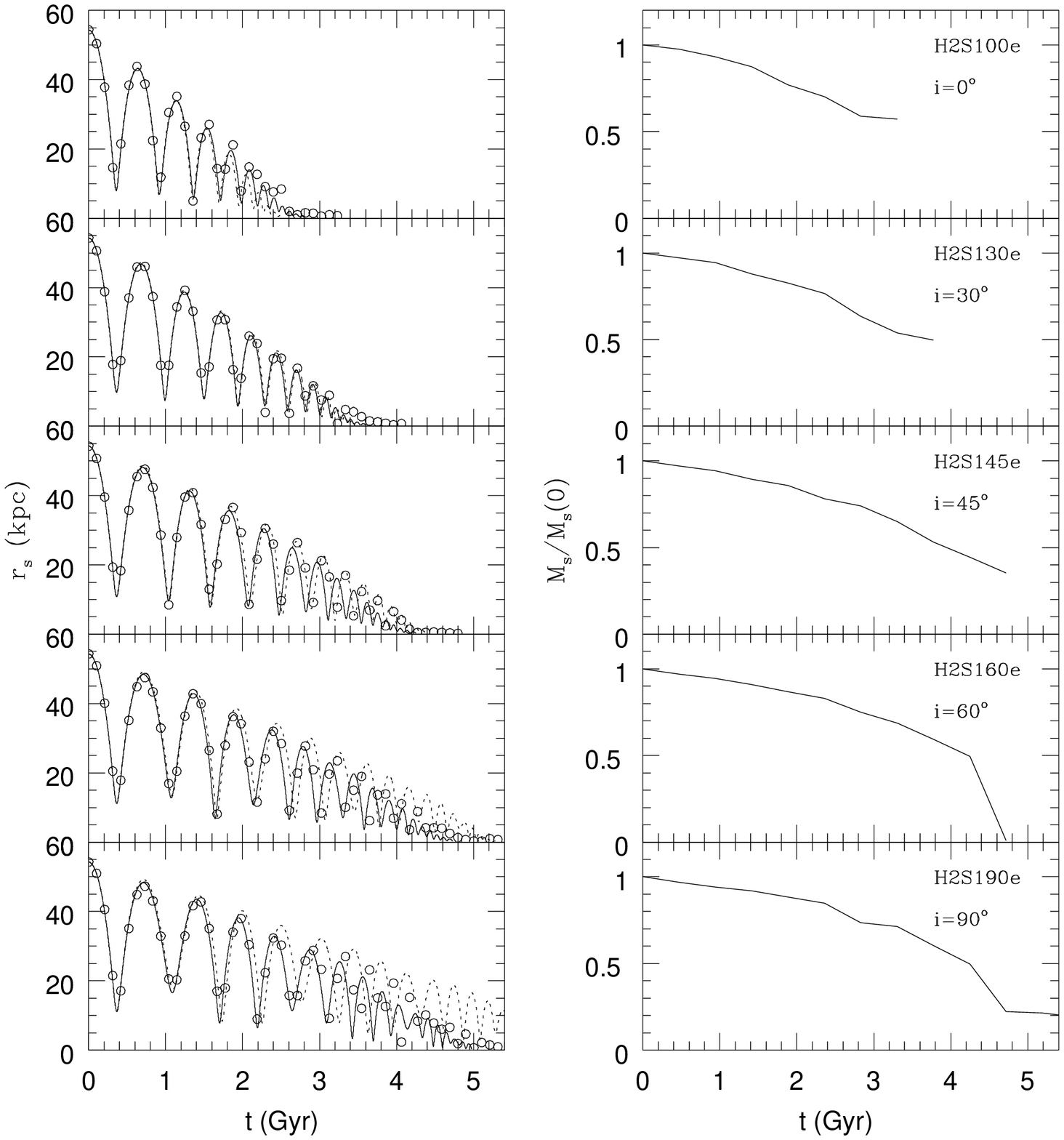} \contcaption{{\bf b:} As
Fig.~\ref{fig:ya1}{\bf a} for those satellites of
Table~\ref{tab:numexp} with initial $e=0.7$. (Note that the time-axis
has changed scale.)}
\end{figure}
\begin{figure}
\vspace{10.5cm} \includegraphics{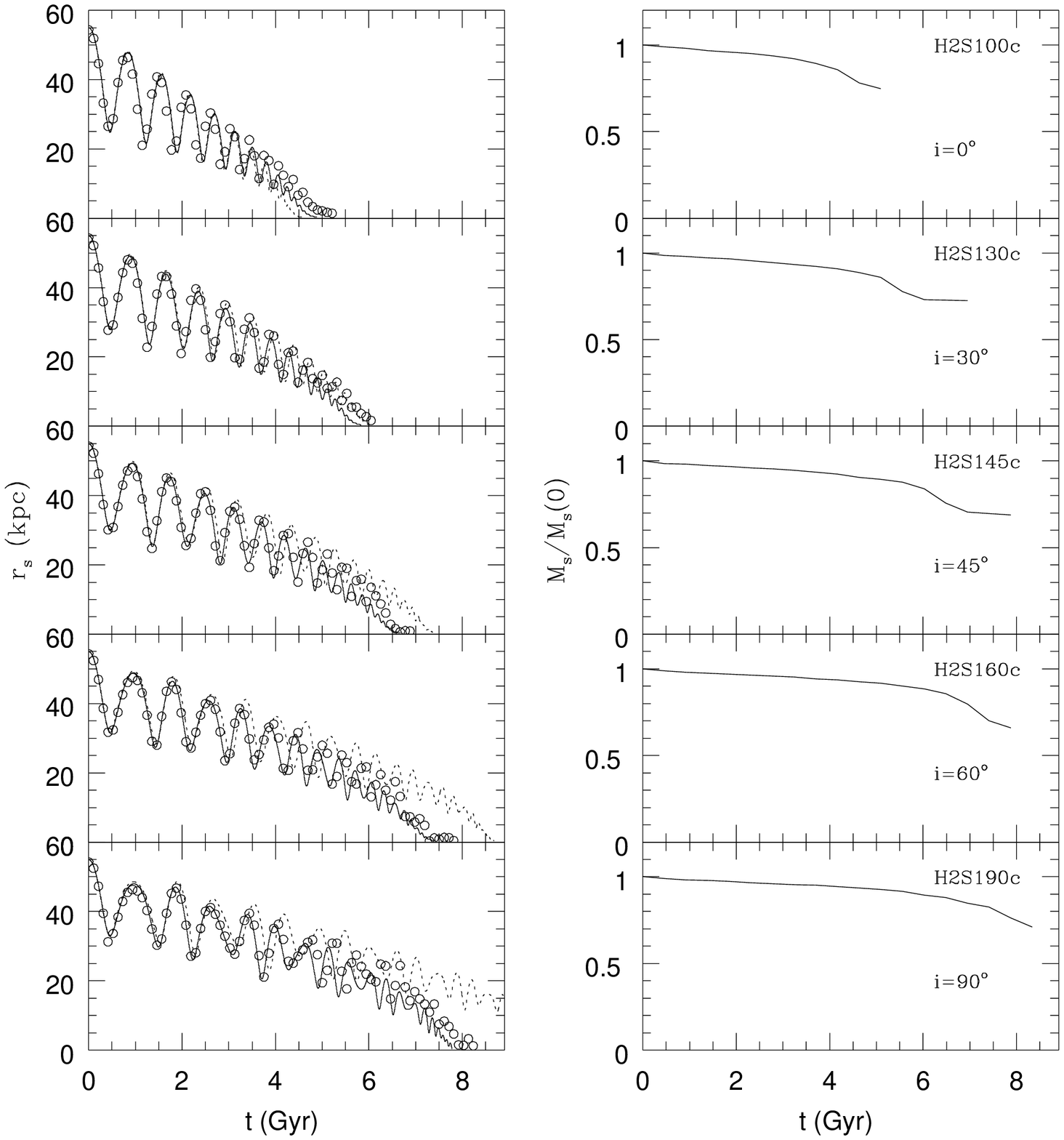} \contcaption{{\bf c:} As
Fig.~\ref{fig:ya1}{\bf a} for those satellites of
Table~\ref{tab:numexp} with initial $e=0.3$. (Note that the time-axis
has changed scale.)}
\end{figure}

PKB observe that coplanar satellites suffer larger friction than those
following polar orbits, leading to survival times 70\% shorter. Due to
the presence of a disc in their galaxy model, the contribution of the
disc anisotropy on the decay differentiation as a function of the
inclination cannot be directly measured. The calculations presented
here (where the disc and bulge are removed) show survival times that
range from 3.7 Gyr (coplanar orbits) to 6 Gyr (polar orbits), using
the same orbital parameters and halo flattening as PKB. This implies a
decay time difference of around 60\% between polar and coplanar
satellites, which indicates that the disc contribution might be of the
order of 10\%. The effects of the disc on the satellite orbit can be found in Pe\~narrubia (2003) and will be addressed in a following paper.

Depending on the symmetry of the halo distribution, one can observe
the following effects:
\begin{itemize}
\item {\bf Spherical mass distribution and isotropic velocity
distribution:} Satellites orbiting systems with a spherical
distribution function move on orbits that do not depend on their
orientation with respect to the symmetry axis.
\item {\bf Flattened mass distribution and isotropic velocity
distribution:} By means of the local approximation the value of
dynamical friction is determined by the properties of the halo at the
satellite's position, being reproduced by Chandrasekhar's formula. Our
results indicate that the spatial asphericity leads to a strong
differentiation of the satellite decay as a function of the orbital
inclination.  
\item {\bf Flattened mass distribution and anisotropic velocity
distribution:} The main influence of the velocity anisotropy on the
satellite orbit is {the secular reduction of the orbital inclination (see
Subsection~\ref{subsec:inclin}) and} the reduction
of the
spatial anisotropy effects, which is equivalent to $B_R < B_z$ in
Binney's formula (eq.~\ref{eqn:ffbinn}, oblate systems). 
{\it Taking into account the velocity anisotropy explicitly thus reduces the
difference in orbital decay times between satellites on polar and
coplanar orbits}.
As
Fig.~\ref{fig:ya1} shows, assuming an isotropic distribution in
velocity space ($B_R=B_z$) or, equivalently, using Chandrasekhar's
formula to reproduce dynamical friction, leads to an overestimation of
dynamical friction for low inclined satellites and to an
underestimation for those satellites following highly inclined orbits.
{This is due to the fact that the anisotropy leads to a reduced effective $X$
for high inclined orbits yielding an enhanced friction force and vice versa for
low inclination. This competes with the density effect due to the flattening, because
 highly inclined
orbits have lower mean density along the orbit compared to lower inclination.}
\end{itemize}

The evolution of the orbital radius of satellites with initial
$e=0.3,0.7$ is plotted in Fig.~\ref{fig:ya1}{\bf b} and {\bf c}. As we
can observe, Binney's approximation reproduces accurately the overall
radius evolution independent of the initial eccentricity and orbital
inclination.

\subsection{Evolution of the orbital inclination and eccentricity}\label{subsec:inclin}
Orbits in non-spherical systems have inclinations $(i\equiv {\rm
arccos}[L_z/L])$ that do not remain constant but suffer periodical
oscillations due to {\em nutation}. In addition, the orbital plane
precesses at a constant rate. Once the initial conditions are fixed,
the amplitude and frequency of nutation and the precession rate remain
constant if the friction force is removed from the equations of motion,
whereas, if implemented, nutation and precession vary according to the
angular momentum and radial distance evolution.  Our interest focuses
now on the effects induced by the velocity anisotropy on the satellite
inclination during the orbit.
\begin{figure}
\vspace{10.5cm}
\includegraphics{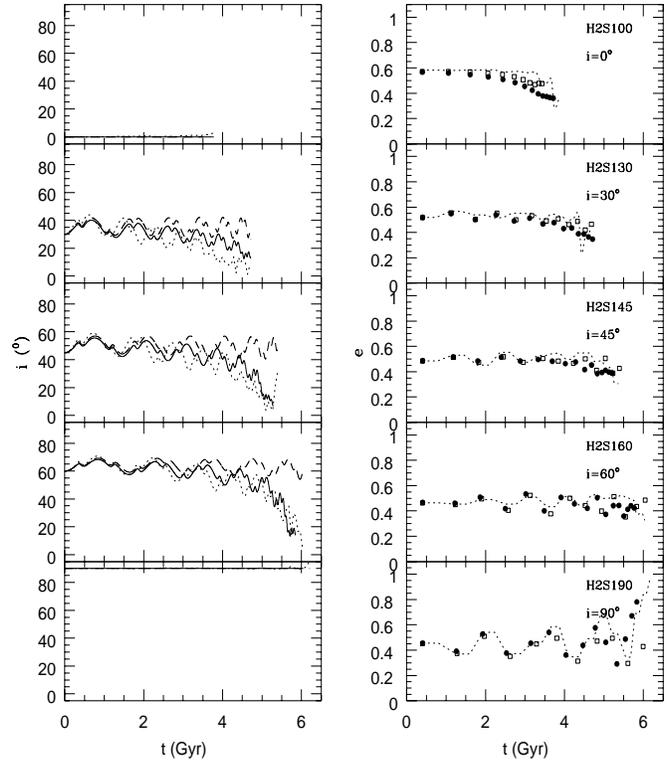}
\caption{{\bf a:} Inclination and eccentricity evolution for the
models of Table~\ref{tab:numexp} with $e=0.5$. Left column: Dotted lines represent
the N-body inclination evolution, full and dashed lines denote the use of Binney's and Chandrasekhar's equations, respectively. Right column: Numerical (dotted lines) against semi-analytical eccentricity evolution. Solid circles and open squares denote the implementation of Binney's and Chandrasekhar's expressions in the semi-analytic code, respectively. We note that, for clarity, the semi-analytic values of $e$ are only plotted at the apo-centres. }
\label{fig:ya4a}
\end{figure}

\begin{figure}
\vspace{10.5cm} \includegraphics{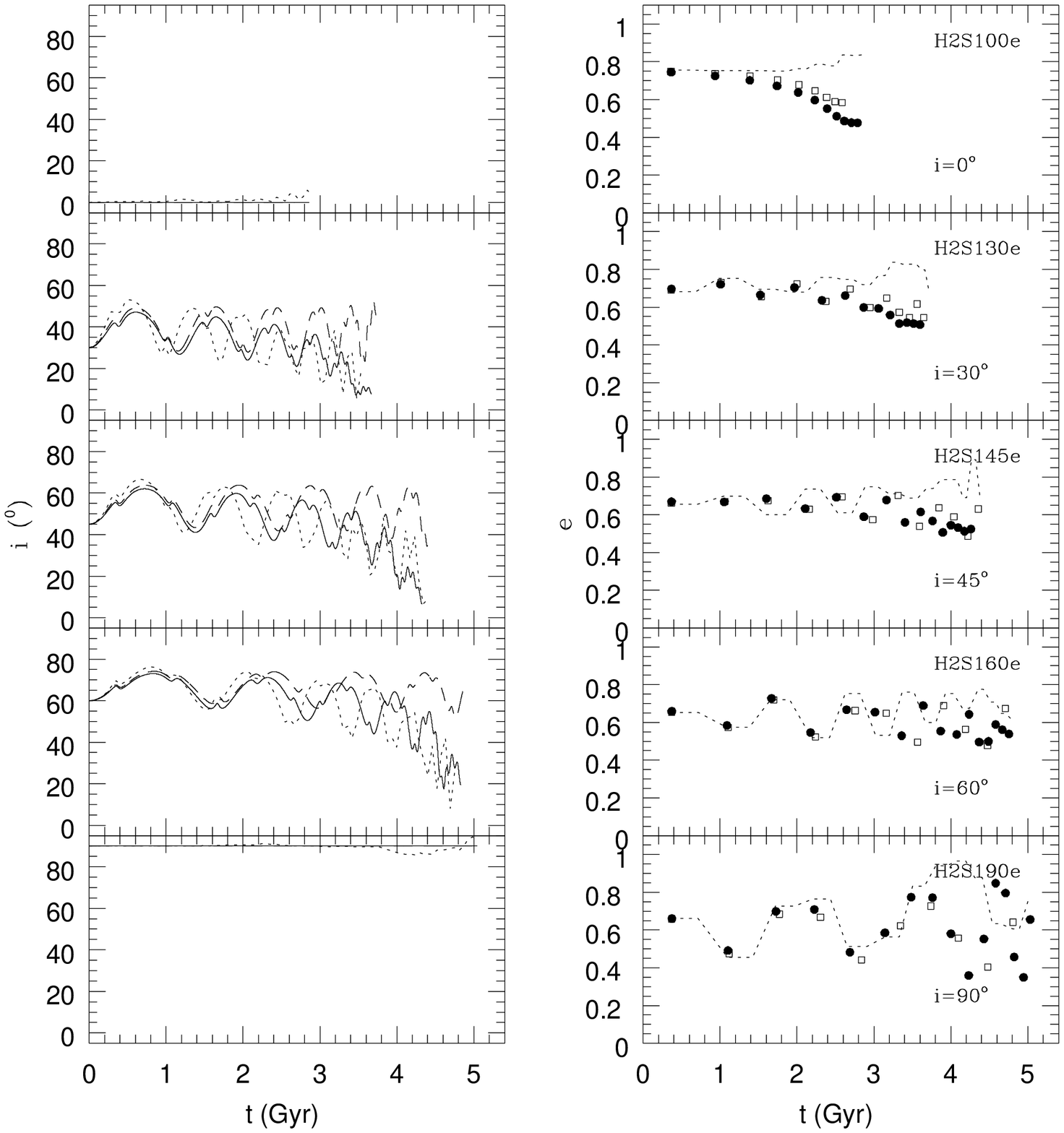} \contcaption{{\bf b:} As Fig.~\ref{fig:ya4a}a
for models with $e=0.7$. Note that the time-scale has a different
value. }
\end{figure}
\begin{figure}
\vspace{10.5cm} \includegraphics{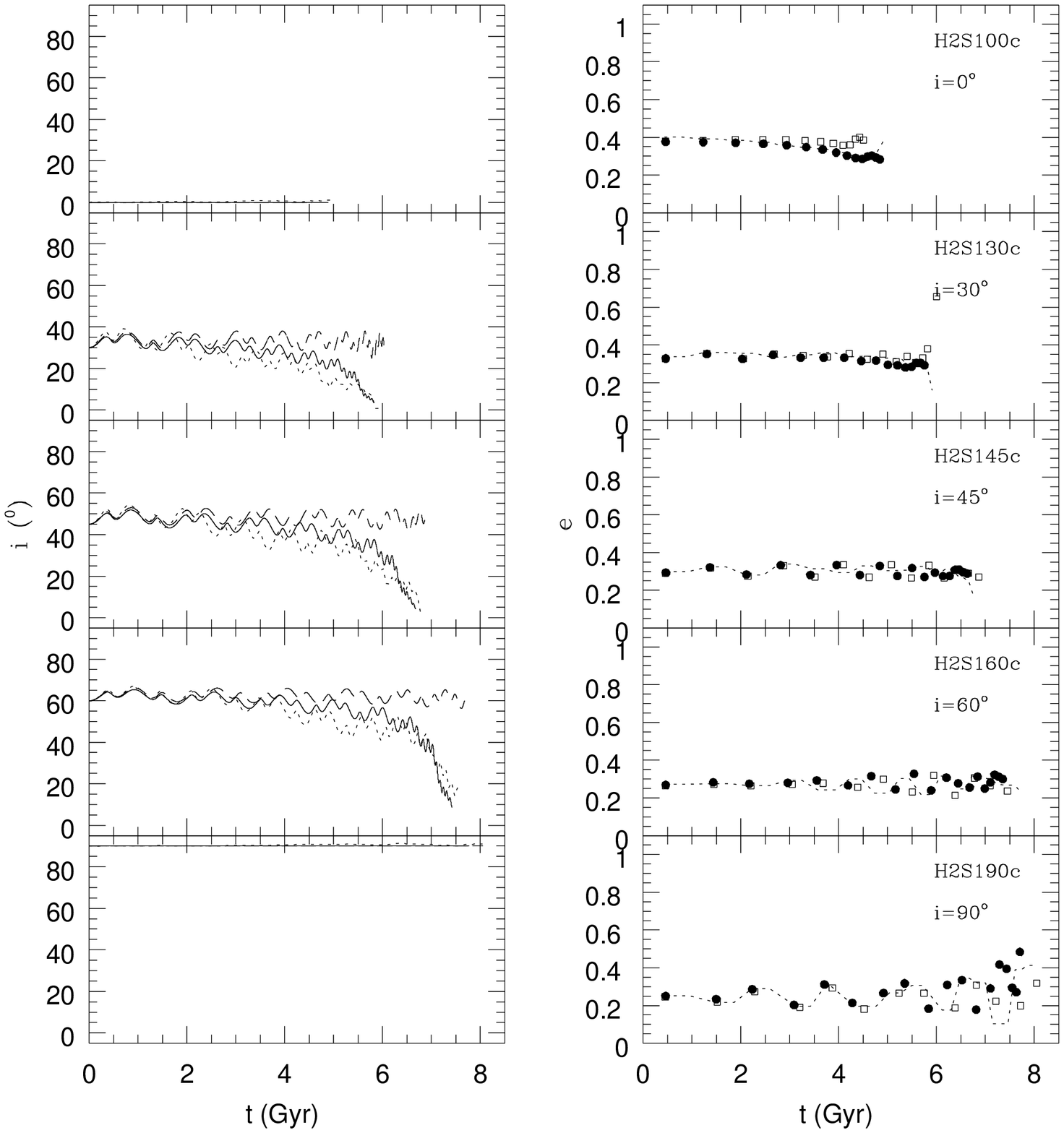} \contcaption{{\bf c:} As Fig.~\ref{fig:ya4a}a
for models with initially $e=0.3$. Note that the time-scale has been
rescaled. }
\end{figure}

Binney (B77) predicts a progressive reduction of $i$ due to dynamical
friction if the velocity dispersion ellipsoid is axi-symmetric
$(\sigma_R,\sigma_z)$ and $\sigma_R>\sigma_z$. By symmetry, the
inclination decrease will not occur if the orbits are either coplanar
$(i=0^\circ)$ or polar $(i=90^\circ)$. 

The inclination evolution of models with $e=0.5$ is plotted in
Fig.~\ref{fig:ya4a} (left column), where dotted lines denote the
numerical data and solid and dashed lines the semi-analytic evolution
if dynamical friction is modelled by Binney's and Chandrasekhar's
formul\ae, respectively. This Figure shows the reduction of the mean
value of $i$ predicted by Binney and observed by PKB in their
numerical experiments. After the satellite has sunk to the inner most
region of the halo, the inclinations are as low as
$10^{\circ}-20^\circ$ independent of their initial value. This
large decrease of $i$ is well reproduced by Binney's expression,
although the nutation process shows discrepancies with the numerical
result, which is connected with the not exact reproduction of the
orbit. Despite the accurate description of the overall decay process,
this orbital mismatch is also observed when applying Chandrasekhar's
expression in spherical systems (see Just \& Pe\~narrubia 2002).  The
figure also confirms that the orbital inclination of coplanar and
polar satellites remains constant.

If dynamical friction is modelled by Chandrasekhar's formula, i.e the
velocity distribution is assumed to be isotropic, the averaged value
of $i$ does not change during the orbit, which 
contradicts the numerical results. 
 Thus, while the difference between the survival times of polar
and coplanar orbits is larger for flattened DMHs treated with
Chandrasekhar's dynamical friction formula (Fig.~\ref{fig:ya1}),
taking into account the DMH anisotropic velocity distribution function
explicitly via Binney's formulae leads to an increase with time of the
anisotropy of the satellite distribution due to the kinematical
coupling of the satellites to the DMH velocity field. This clearly agrees with the fully self-consistent $N-$body computations
reported here.

In Fig.~\ref{fig:ya4a}{\bf b} and ~\ref{fig:ya4a}{\bf c} (left columns)
we plot the comparison for models with $e=0.7, 0.3$, respectively. The
results show barely a dependence on the eccentricity. It is
interesting to note that, independently of $e$, orbits that are
neither coplanar nor polar present large drops of the mean value of
$i$. After the satellite sinks to the centre, the final orbital
inclination lies in between 10-20$^\circ$ for all the models.

We must emphasise the accuracy of Binney's formula in describing
correctly the inclination decrease that satellites suffer in oblate
systems. This is crucial for simulating properly satellite motions and
for investigating satellite distributions around spiral galaxies.

Like the orbital inclination, the eccentricity is one of the orbital
parameters that can be indirectly measured from observations to
determine a satellite's motion around a galaxy. In the right columns of
Fig.~\ref{fig:ya4a} we show the comparison of the numerical
eccentricity evolution with both semi-analytic approaches. The
analytic formul\ae~of dynamical friction lead to a larger eccentricity
decrease, which occurs mostly at late-times of the orbit, the
so-called {\em orbital circularisation}, and becomes stronger for low
inclined orbits, those that suffer higher dynamical friction.
Fig.~\ref{fig:ya4a}{\bf b} and Fig.~\ref{fig:ya4a}{\bf c} indicate that
the circularisation is more pronounced if the initial orbital
eccentricity is higher and decreases for more circular orbits.  Both
dynamical friction expressions lead to a similar eccentricity
evolution for the first few orbital periods, however at late-times the
eccentricity exhibits a reduction not present in the numerical
calculations that can be as high as $\sim 80\%$ for low inclined
satellites on highly eccentric orbits (Fig.~\ref{fig:ya4a}{\bf b},
model H2S100e).

\subsection{Energy and angular momentum evolution}
A flattened system possesses two analytic constants of motion, the
energy and the component of the angular momentum perpendicular to the
axi-symmetry plane (which we denote as $L_z$). The total angular
momentum $L^2=L_R^2+L_z^2$ is, however, not constant during a
satellite orbit (see e.g Binney \& Tremaine 1987), but has periodic
variations that correspond to a precession and nutation of the orbital
plane around the $z$-axis.

Since the dynamical friction force has an opposite sense with respect
to the satellite velocity, it decreases the angular momentum and the
energy which induces a monotonic sinking into the inner regions of the
halo potential. The reduction of angular momentum, therefore, implies
an increase of the binding energy (in absolute value), since the
potential grows with decreasing radius.  Due to the small magnitude of
dynamical friction compared to the mean field force, we expect an
easier comparison between numerical and semi-analytic data by the slow
variation of $L_z$ and $E$ during the orbit.
\begin{figure}
\vspace{10.5cm} \includegraphics{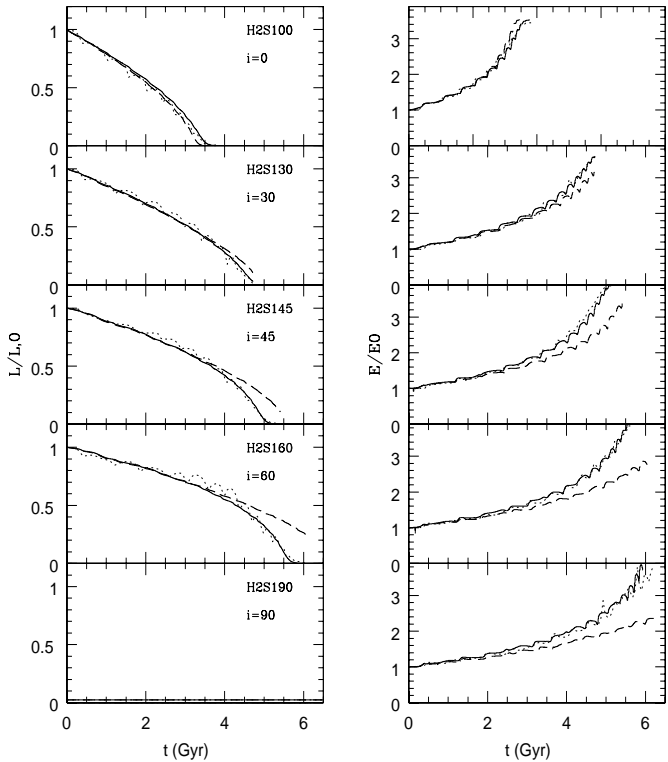}
\caption{ {Specific} energy and angular momentum evolution during the orbits with
$e=0.5$. The numerical evolution is denoted by dotted lines, whereas
the semi-analytic data is represented by solid and dashed lines if
dynamical friction is modelled by Binney's and Chandrasekhar's
formul\ae, respectively. The quantities $E$ and $L_z$ are normalised
to the initial value. Note that for the case $i=90^\circ$ one has $L_z=0$.}
\label{fig:ya7}
\end{figure}

In Fig.~\ref{fig:ya7} we plot the changes of {specific} $E$ and $L_z$ due to
dynamical friction for the models with $e=0.5$. The results are
equivalent to those of the radial evolution. Due to our selection of
$\ln \Lambda$ as the average over those Coulomb logarithms that lead
to the best fit for each particular model, Chandrasekhar's formula
overestimates dynamical friction for low inclined orbits and
underestimates it for highly inclined orbits. For orbits with $i<
30^\circ$, this appears as a stronger reduction of the $z$-component
of angular momentum and, equivalently, a large increase of the {binding}
energy. The effect is contrary for satellites with $i>30^\circ$.  

This figure illustrates how the kinetic energy of the satellite is
lost via friction, being {taken-up partially by halo particles and also being deposited in 
unbound satellite particles}. At the end of the simulation
the angular momentum has a null value, i.e the satellite remains in
the inner most part of the galaxy.

It is interesting to note that the numerical evolution of energy and
angular momentum presents small oscillations during the orbit. This
behaviour is due to the self-response of the halo to the satellite
motion. Since {\sc superbox} preserves the total energy and angular
momentum, the halo also moves around the centre-of-mass of the
system. Due to the complexity of the feedback, it cannot be reproduced
analytically (the halo centre-of-mass is fixed at the coordinate
origin in the semi-analytic code).

\section{Conclusions}\label{sec:concluflat}
To asses the accuracy of Binney's equations (B77) and in order to
reproduce the decay of satellites in flattened DMHs with a
semi-analytical code, we have performed a set of self-consistent
numerical experiments for different orbital inclinations and
eccentricities of the satellite.  The semi-analytic code incorporates
dynamical friction either in terms of Chandrasekhar's expression or as
Binney's formulae.  Both
treatments include the aspherical density profile by means of the local approximation. This means that the
differences on the satellite motion induced by each treatment of
dynamical friction comes from the anisotropy in velocity space, which
is implemented in the analysis of B77.

The accuracy of Binney's and Chandrasekhar's formul\ae~in comparison
with the numerical orbits is determined by the parameter $\chi^2=\frac{1}{2k}
\sum(\Delta {\bf r}^2+\sigma^2\Delta t^2)$  at the peri- and apo-centres for
a given number $k$ of orbits. If dynamical friction is modelled by
Binney's equation, this quantity shows discrepancies of approximately
$\chi_{\rm min}=0.009 r_0$ per orbit after averaging over the set of
experiments and for the first three orbits, while Chandrasekhar's
formula produces values of around $\chi=0.05 r_0$ per orbit (Table~\ref{tab:averres}).

We conclude that Binney's expression faithfully reproduces the process
of dynamical friction in anisotropic systems. The fit is as accurate
as that employing Chandrasekhar's formula in isotropic systems (see
Just \& Pe\~narrubia 2002).

The comparison of orbits resulting from Chandrasekhar's and Binney's
expression of dynamical friction gives us the possibility to asses the
effects of the DMH velocity anisotropy on satellite dynamics. We have
demonstrated that, (i) if the density profile is in both equations
$\rho=\rho(R,z)$, where $R,z$ are the cylindrical coordinates of the
satellite position vector, the orbits generated by Chandrasekhar's
formula overestimate the decay time for polar orbits and underestimate
it for coplanar ones for an overall best-fit Coulomb logarithm
(ln$\Lambda=2.2$).  One effect of the velocity anisotropy is then to
reduce the interval of decay times as a function of the orbital
inclination. Binney's expression has been shown to reproduce
accurately the numerical results independently of the initial
eccentricity and with ln$\Lambda=2.4$.  (ii) Dynamical friction in
systems with anisotropic velocity distribution leads to a marked
decrease of the orbital inclination $i$ which is well reproduced by
Binney's expression. After the satellite sinks to the inner most
region of the galaxy, $i$ lies within 10-20$^\circ$, independently of
the initial value unless $i\approx90^{\rm o}$ (polar orbits). (iii) The energy and angular momentum evolution as
a function of the orbital inclination confirm the results of (i) and
(ii).

The semi-analytic eccentricity evolution, either employing
Chandrasekhar's or Binney's formula, shows the so-called {\em
circularisation} process, defined as the progressive reduction of $e$
during the orbit. This variation is stronger for increasing friction
(as for coplanar orbits or during the late-times of the evolution) and
barely takes place in the self-consistent $N-$body calculations. The small $N-$body circularisation
agrees with the results of van den Bosch et al. (1999). 
 A possible solution
may be sought in the position-dependence of the Coulomb logarithm, as
proposed by Hashimoto, Funato \& Makino (2003). However, despite the
improvement in the description of the orbit at early-times, the scheme
of Hashimoto et al. 
overestimates the satellite decay time for all the experiments
(see Just \& Pe\~narrubia 2002).  
 We must conclude that the reason for circularisation in the semi-analytic orbits is not yet fully understood.

\section{Acknowledgements} 
JP acknowledges support through a SFB~439 grant at the University of
Heidelberg. PK acknowledges support through DFG grant KRI635/4. We thank the anonymous referee for his/her advice and useful comments.

{}

\end{document}